\documentclass{article}
\usepackage{amssymb}
\usepackage{amsfonts}
\usepackage{amsmath}
\usepackage{graphicx}

\input{tcilatex}

\begin{document}

\title{Space-time anisotropy: theoretical issues and the possibility of an
observational test. }
\author{Sergey Siparov \\
State University of civil aviation, St-Petersburg, Russia\\
sergey@siparov.ru \and Nicoleta Brinzei \\
''Transilvania'' University, Brasov, Romania\\
nico.brinzei@rdslink.ro}
\maketitle

\begin{abstract}
The specific astrophysical data collected during the last decade causes the
need for the modification of the expression for the Einstein-Hilbert action,
and several attempts sufficing this need are known. The modification
suggested in this paper stems from the possible anisotropy of space-time and
this means the natural change of the simplest scalar in the least action
principle. To provide the testable support to this idea, the optic-metrical
parametric resonance is regarded - an experiment on the galactic scale based
on the interaction between the electromagnetic radiation of cosmic masers
and periodical gravitational waves emitted by close double systems or
pulsars. Since the effect depends on the space-time metric, the possible
anisotropy could reveal itself through observations. To give the
corresponding theory predicting the corrections to the expected results of
the experiment, the specific mathematical formalism of Finsler geometry was
chosen. It was found that in case the anisotropy of the space-time exists,
the orientation of the astrophysical systems suitable for observations would
show it. In the obtained geodesics equation there is a direction dependent
term.
\end{abstract}

\section{Introduction}

The amount of astrophysical data collected during the last decade and
contradicting the mainstream solutions of general relativity makes one think
that the least action principle based on the \textquotedblright simplest
scalar\textquotedblright\ in the form of Ricci scalar curvature does not
work properly in some important cases (e.g. rotation curves for spiral
galaxies \cite{begeman}). The efforts to improve the situation were
undertaken in the following directions: complicating of the existing scalar
(including $f(R)$ theories); change of the scalar (for example, the use of
Weyl tensors); introduction of additional scalar fields (including MOND
theory); passing to a non-symmetrical metric.

An unusual, but fundamental assumption of the anisotropy of the space-time
means that in this case its geometry must be not Riemannian, but Finslerian.
It means that the \textquotedblright simplest scalar\textquotedblright\ will
become more complicated in a natural way. Obviously, this brings a lot of
consequences which means that, first of all, we need a way to test the
validity of this theoretical approach for real physics. The corresponding
test can be performed with the help of observations based on the effect of
the optic-metrical parametric resonance (OMPR). In this paper we give the
brief theory of the OMPR effect in order to understand what kind of
corrections due to anisotropy might be of theoretical and experimental
interest. After that we introduce the mathematical formalism needed to
construct simple models of the anisotropic space and then regard two types
of metrics. Every time we present the modifications of the OMPR conditions
that could make it possible to discover the space-time anisotropy (if any)
on the galactic scale. In the Discussion the main results are given.

\section{Brief theory of the OMPR effect}

Let us regard a two-level atom in the strong monochromatic quasi-resonant
field. The system of Bloch's equations for the components of the density
matrix components is

\begin{align}
\frac d{dt}\rho_{22} & =-\gamma\rho_{22}+2i\alpha_{1}\cos(\Omega
t-ky)(\rho_{21}-\rho_{12})  \label{eq.1} \\
(\frac\partial{\partial t}+v\frac\partial{\partial y})\rho_{12} &
=-(\gamma_{12}+i\omega)\rho_{12}-2i\alpha_{1}\cos(\Omega
t-ky)(\rho_{22}-\rho_{11})  \notag \\
\rho_{22}+\rho_{11} & =1  \notag
\end{align}

Here $\rho_{22}$ and $\rho_{11}$ are the populations of the levels, $\rho
_{12}$ and $\rho_{21}$ are the polarization terms, $\gamma\ $and $%
\gamma_{12} $ are the longitudinal and transversal decay rates of the atom
(if level $1$ is the ground level, $\gamma_{12}=\gamma/2$); $\alpha_{1}=%
\frac{\mu E}\hbar$ is the Rabi parameter (Rabi frequency) proportional to
the intensity of the electromagnetic wave (EMW), $\mu$ is the dipole
momentum, $E$ is the electric stress, $k$ is the $y$-component of the wave
vector of the EMW, $v$ is the atom velocity along the $Oy$-axis pointing at
the detector, $\gamma <<\alpha_{1}$ is the condition of the strong field.
The dynamics of this system in cases when various parametric resonances are
possible was investigated in\cite{sip-prev}.

Let this atom belong to a saturated space maser in the field of the periodic
gravitational wave (GW) emitted by a compact binary star or by a pulsar and
propagating anti-parallel to the $Ox$-axis pointing at the GW-source. The GW
acts on the atomic levels, on the maser radiation and on the geometrical
location of the atom. In \cite{sip-A} it was shown that the first effect is
much smaller than the other two effects. The action of the GW on the
monochromatic EMW could be accounted for by the solution of the eikonal
equation

\begin{equation}
g^{ik}\frac{\partial\psi}{\partial x^{i}}\frac{\partial\psi}{\partial x^{k}}%
=0;~\ \ i,k=1\div4  \label{eq.2}
\end{equation}

The atom velocity, $v$, could be obtained from the solution of the geodesic
equation

\begin{equation}
\frac{d^{2}x^{i}}{ds^{2}}+\Gamma_{kl}^{i}\frac{dx^{k}}{ds}\frac{dx^{l}}{ds}%
=0;~\ \ i,k,l=1\div4  \label{eq.3}
\end{equation}
(and not from the solution of the geodesic declination equation as in the
calculations of the displacements of the parts of the laboratory setups,
designed for the detection of the GW).

The equations (\ref{eq.1}-\ref{eq.3}) are basic for the theory of the OMPR
effect. Such a signal being registered could provide the possibility to
detect the GW in a principially new way which differs from the other 18
known ones \cite{glad} by the fact that the OMPR effect is of the zero order
and not of the first order in the non-dimensional amplitude of the GW.

If we use the regular Riemann geometry, the solution gives the following.
The weak gravitational field in the empty space (far from masses) is
described by the linearized Einstein equations. Then the perturbation $%
h^{k}{}_{i}$ of the flat space metric tensor (Minkowski metric) suffices the
wave equation, which has the solution \cite{amaldi}

\begin{equation}
h^{k}{}_{i}=Re[A^{k}{}_{i}\exp(iK_{\alpha}x^{\alpha})]  \label{eq.5}
\end{equation}
where $K^{\alpha}$ is a light-like\emph{\ }vector, i.e.it satisfies the
equation $K_{\alpha}K^{\alpha}=0$. Then the perturbed metric tensor in the
isotropic case can be written as

\begin{equation}
g^{ik}=\left( 
\begin{array}{cccc}
1 & 0 & 0 & 0 \\ 
0 & -1 & 0 & 0 \\ 
0 & 0 & -1+h\cos\frac{D}{c}(x^{1}-x^{2}) & 0 \\ 
0 & 0 & 0 & -1-h\cos\frac{D}{c}(x^{1}-x^{2})%
\end{array}
\right)  \label{eq.6}
\end{equation}
where $h$ is the dimensionless (small) amplitude of the GW, $D$ is the
frequency of the GW, $x^{i},(i=1\div4)$ are the coordinates.

The solution of (\ref{eq.2}) with regard to (\ref{eq.6})

\begin{equation}
\psi=\frac{\omega}{c}x^{1}+\sum_{i=2}^{4}k_{i}x^{i}-hc^{2}\frac{%
k_{3}^{2}-k_{4}^{2}}{4\omega D}\sin\frac{D}{c}(x^{1}-x^{2})  \label{eik-R}
\end{equation}
shows that the action of the GW causes \emph{a} phase modulation of the EMW.
Since $h$ is very small, the phase modulated EMW can be presented (in
Cartesian coordinates) as a superposition \cite{sip-L6}

\begin{equation}
E(t)=E\cos(\Omega t-ky)+E\frac\omega{8D}h[\cos((\Omega-D)t-ky)-\cos
((\Omega+D)t-ky)]  \label{eq.7}
\end{equation}

The solution of (\ref{eq.3}) with regard to (\ref{eq.6}) gives \cite{sip-A}

\begin{equation}
y(t)\sim h\frac cD\sin(Dt+Kx)  \label{eq.8}
\end{equation}
where $K$ is the GW wave vector. The expression (\ref{eq.8}) makes it
possible to get the component of the atom velocity directed towards the Earth

\begin{align}
v & =v_{0}+v_{1}\cos Dt  \label{eq.9} \\
v_{1} & =hc  \notag
\end{align}

Substituting (\ref{eq.9}) and (\ref{eq.7}) into (\ref{eq.1}), one gets

\begin{align}
\frac d{dt}\rho_{22} & =-\gamma\rho_{22}+2i[\alpha_{1}\cos(\Omega
t-ky)+\alpha_{2}\cos((\Omega-D)t-ky)  \label{eq.10} \\
& \ -\alpha_{2}\cos((\Omega+D)t-ky)](\rho_{21}-\rho_{12}) \\
\frac d{dt}\rho_{12} &
=-(\gamma_{12}+i\omega)\rho_{12}-2i[\alpha_{1}\cos(\Omega
t-ky)+\alpha_{2}\cos((\Omega-D)t-ky)  \notag \\
& \ -\alpha_{2}\cos((\Omega+D)t-ky)](\rho_{22}-\rho_{11}) \\
\rho_{22}+\rho_{11} & =1  \notag
\end{align}
where $\alpha_{2}=\frac{\omega h}{8D}\alpha_{1}$, and (\ref{eq.9}) was used
in the the expression for the full time derivative $\frac d{dt}=\frac 
\partial{\partial t}+kv$. The solution of the system (\ref{eq.10}) is
performed by the asymptotical expansion method, the small parameter being $%
\varepsilon=\frac\gamma{\alpha_{1}}$ (notice, that $\frac{\alpha_{2}}{%
\alpha_{1}}\sim\varepsilon$ too). The essential point is the OMPR conditions

\begin{align}
\frac\gamma{\alpha_{1}} & =\Gamma\varepsilon;\Gamma=O(1);\varepsilon <<1
\label{eq.11} \\
\frac{\alpha_{2}}{\alpha_{1}} & =\frac{\omega h}{8D}=b\varepsilon
;b=O(1);\varepsilon<<1 \\
\frac{kv_{1}}{\alpha_{1}} & =\frac{\omega h}{\alpha_{1}}=\kappa
\varepsilon;\kappa=O(1);\varepsilon<<1 \\
(\omega-\Omega+kv_{0})^{2}+4\alpha_{1}^{2} & =D^{2}+O(\varepsilon
)\Rightarrow D\sim2\alpha_{1}
\end{align}
If they are fulfilled, then the principal term of the asymptotic expansion
for $Im(\rho_{21})$ which characterizes the scattered radiation energy flow
can be calculated explicitly. The effect of the OMPR is that at a frequency
shifted by $D$ from the central peak of the EMW (that is from the signal of
the space maser), the energy flow is proportional to $\varepsilon^{0}$, i.e.
has zero order in the powers of the small parameter of the expansion, and
has the form

\begin{equation}
Im(\rho_{21})\sim\frac{\alpha_{1}}D\cos2Dt+O(\varepsilon)  \label{eq.12}
\end{equation}

It means that the energy flow at this frequency is periodically amplified
and attenuated with the (doubled) frequency of the GW. The OMPR signal can
be registered with the help of the special statistical processing of the
radio telescope signal. The registration of such a signal would give a new
experimental evidence of the GW existence and since the value of the OMPR
signal is comparable to that of the regular maser peak, such observations
could lead to the design of a GW map of the sky.

Turning back to the problems mentioned in the Introduction, one can suggest
that such kind of measurements could be able to give an evidence of the
space-time anisotropy (if any) on the galactic scale - that is on the scale
where problems appear. In this case the use of the Finslerian
Hilbert-Einstein action instead of the Riemannian one would be grounded. In
order to develop a theory supporting such forthcoming observational results,
we have first of all to show explicitly that Einstein equations in empty
space for the anisotropic case still have the form of the wave equation
though its solution might become dependent on the direction. (In the
qualitative analysis presented in \cite{sip-C6}, \cite{sip-book}, \cite%
{sip-M7} it was presumed so). Then we have to perform the corresponding
modifications of the eikonal and geodesics equations and use the results to
describe the changes in the OMPR signal.

\section{Mathematical formalism and basic equations}

Let $M=\mathbf{R}^{4}$\ be regarded as a differentiable 4-dimensional
manifold of class $C^{\infty},$\ $TM$\ its tangent bundle, and $%
(x,y)=(x^{i},y^{i});$\ $i=1,..,4$ the coordinates in a local frame. We call 
\textit{locally Minkowskian} a metric with the property that there exists a
system of local coordinates on $TM\ $in which it\emph{\ }does not depend on
the positional variables, $x^{i}$, but may depend on the directional
variables, $y^{i}=\dfrac{\partial x^{i}}{\partial t}$ ($t$ is a parameter), $%
i=1\div4$. Let us regard a metric tensor $g_{ij}(x,y)=\gamma_{ij}(y)+%
\varepsilon_{ij}(x,y),$ $\forall(x,y)\in TM$, where $\gamma=\gamma(y)$ is a
Finsler-locally Minkowskian undeformed\textit{\ }metric tensor on $M$ and%
\textit{\ }$\varepsilon=\varepsilon(x,y)$ is a small anisotropic deformation
of $\gamma$. In anisotropic spaces, the tangent spaces $T_{x}M,$ $x\in M$
are generally curved. The general approach\emph{\ }we shall use\emph{\ }was
developed by R. Miron and M. Anastasiei \cite{Lagrange}, \cite{Miron} and is
known as $h$-$v$ metric model formalism. Some specific models close to that
under discussion here were given in \cite{Mangalia}\ and \cite{BB}. In these
models, $TM$ turns to be a Riemannian manifold of dimension $8.$

The Ricci scalar $\mathcal{R=}$ $G^{\alpha\beta}\mathcal{R}_{\alpha\beta} $
used in the Einstein-Hilbert action in the case of $h$-$v$ models \cite%
{Lagrange}, \cite{BB} leads to the expression $\mathcal{R}=R+S,$ where $S$
is the Ricci scalar of the tangent (Riemannian) space $T_{x}M,$ for $x\in M.$
In our approach, we shall choose for simplicity a model in which $S=0$ while 
$R$ depends on the direction, $y,$ through the $y$-dependence of the
Christoffel symbols, $\Gamma_{~jk}^{i},$ calculated with regard to $g(x,y),$
that is $\Gamma_{~jk}^{i}=\dfrac12\gamma^{ih}(\dfrac{\partial\varepsilon_{hj}%
}{\partial x^{k}}+\dfrac{\partial\varepsilon_{hk}}{\partial x^{j}}-\dfrac{%
\partial\varepsilon_{jk}}{\partial x^{h}})$. The only components of the
Ricci tensor $\mathcal{R}_{\alpha\beta}=(R_{jk},\overset{1}{P}\overset{}{%
_{bj}},$ $\overset{2}{P}\overset{}{_{jb}},S_{ab})$ that do not identically
vanish for our model are $R_{jk}=\dfrac{\partial\Gamma_{~jk}^{i}}{\partial
x^{i}}-\dfrac{\partial\Gamma_{~ji}^{i}}{\partial x^{k}}+\Gamma_{~jk}^{h}%
\Gamma_{~hi}^{i}-\Gamma_{~ji}^{h}\Gamma_{~hk}^{i}$ and the ''mixed''
component $\overset{2}{P}\overset{}{_{jb}}=P_{j~ib}^{~i}=\dfrac12(%
\delta_{s}^{i}\delta_{j}^{l}-\gamma^{il}\gamma_{sj})\dfrac{%
\partial\Gamma_{~li}^{s}}{\partial y^{b}}. $ Hence, the Ricci scalar in this
simple model turns to be $R=\gamma^{jk}R_{~jk}.$

\subsection{Einstein equations}

The Einstein equations for the empty space \cite{Lagrange}, \cite{Miron}, 
\cite{BB} appear to be, for our linearized model:

\begin{align}
R_{ij}-\dfrac12R\gamma_{ij} & =0  \label{E1} \\
(\delta_{s}^{i}\delta_{j}^{l}-\gamma^{il}\gamma_{sj})\frac{\partial
\Gamma_{~li}^{s}}{\partial y^{b}} & =0  \label{E2}
\end{align}

The first set of equations (\ref{E1}) involves only the $x$-derivatives of
the deformation, $\varepsilon,$ while the second ones, (\ref{E2}), contain
mixed derivatives of $\varepsilon$ of the second order. In order to
integrate the first equations (\ref{E1}), we apply the same procedure as in
the classical Riemannian case and look for the solutions satisfying the
harmonic (Lorentz) gauge conditions\textit{\ }$\gamma^{ij}%
\Gamma_{~ij}^{h}=0, $ which are actually 
\begin{equation}
\frac{\partial\varepsilon_{j}^{i}}{\partial x^{i}}-\dfrac{1}{2}\frac {%
\partial\varepsilon}{\partial x^{j}}=0.  \label{har}
\end{equation}

Consequently, the first set of equations (\ref{E1}) becomes 
\begin{equation}
\square\varepsilon_{ij}=0;  \label{waveq}
\end{equation}
which demonstrates explicitly the existence of the GW in the anisotropic
space. We look for a wave solution in the form 
\begin{equation}
\varepsilon_{jh}=\func{Re}(a_{jh}(y)e^{iK_{m}(y)x^{m}}),  \label{wavesol}
\end{equation}
where $i$ denotes the imaginary unit. Strictly speaking, we should take the
perturbation as a series each term of which corresponds to a wave. But for
simplicity, we regard only one term of this series. Both the amplitude $%
a_{jh}(y)$ and the wave vector $K(y)$ of the GW are no longer isotropic, but
may\emph{\ }depend on direction. Substituting \ref{wavesol} into \ref{waveq}%
, we see that either $\varepsilon$ itself is zero, or we must have $\gamma
^{hl}K_{h}K_{l}=0.$ By (\ref{E1}), and (\ref{har}), we infer that in the
anisotropic space $a_{jh}(y)$ and $K_{m}(y)$ should obey the algebraic
system 
\begin{equation}
\left\{ 
\begin{array}{c}
\gamma^{hl}K_{h}K_{l}=0 \\ 
a_{~j}^{i}K_{i}=\dfrac{1}{2}a_{~i}^{i}K_{j}%
\end{array}
\right.  \label{conds}
\end{equation}

Thus, the wave solutions (\ref{wavesol}) obeying (\ref{har}) of the Einstein
equations (\ref{E1}, \ref{E2}) must suffice: 
\begin{equation}
\left\{ 
\begin{array}{l}
\gamma^{hl}K_{h}K_{l}=0 \\ 
a_{~j}^{i}K_{i}=\dfrac12a_{~i}^{i}K_{j} \\ 
\frac\partial{\partial y^{b}}(\dfrac12a_{~i}^{i}K_{j})=\overset{0}{C}\overset%
{}{_{~lb}^{i}}(2a_{~j}^{l}K_{i}-a_{~i}^{l}K_{j}).%
\end{array}
\right.  \label{wavesol2}
\end{equation}
where the third equation comes from the \textquotedblright mixed Einstein
equation\textquotedblright\ (\ref{E2}), $\overset{0}{C}\overset{}{_{~lb}^{i}}%
=\dfrac12\gamma^{ih}\dfrac{\partial\gamma_{hl}}{\partial y^{b}}.$ We also
see that the amplitude $a_{~j}^{i}$ and the wave vector $K_{i}$\ now depend
on each other.

Such solutions will behave tensorially under coordinate transformations $%
\tilde x^{i}=\Lambda_{j}^{i}x^{j}$ ($\Lambda_{j}^{i}\in\mathbf{R})$ on the
manifold $M=\mathbf{R}^{4}$ (which include homogeneous Lorentz
transformations, \cite{Carroll}, in the Minkowski case and the group $%
G_{1}(P_{k+2m}),$ \cite{Pavlov-Garasko}, in the more complicated
Berwald-Moor case to be regarded below).

\subsection{Eikonal equation}

In case when the space is anisotropic, the first approximation of the
eikonal, $\psi(x^{i},y^{i}),$ corresponding to some wave (e.g. to the EMW),
can be written as $\psi=\psi_{0}+\dfrac{\delta\psi}{\delta x^{i}}dx^{i}+%
\dfrac{\partial\psi}{\partial y^{a}}\delta y^{a}$. Then, (\ref{eq.2})
becomes: 
\begin{equation}
g^{ij}\dfrac{\delta\psi}{\delta x^{i}}\dfrac{\delta\psi}{\delta x^{j}}=0,
\label{h_eiko}
\end{equation}
where the ''adapted derivative'' $\dfrac\delta{\delta x^{i}}$ is
characteristic for Finsler and Lagrange geometries, \cite{Miron}, \cite%
{Lagrange} and insures the tensorial character of $\dfrac{\delta\psi }{%
\delta x^{i}}.$ Under our assumptions on the metric structure and on
coordinate transformations, this equation becomes simply: 
\begin{equation}
g^{ij}\dfrac{\partial\psi}{\partial x^{i}}\dfrac{\partial\psi}{\partial x^{j}%
}=0.  \label{simple_eiko}
\end{equation}

Let us look for the eikonal in the form $\psi(x,y)=\widehat{f}(x,y)+h%
\widehat {g}(x,y);$ ($h<<1,h^{2}\simeq0)$ that satisfies (\ref{simple_eiko}%
). With $g^{ij}=\gamma^{ij}-h\tilde a^{ij}\cos(K_{m}x^{m}),$ for our model,
we get 
\begin{equation}
\gamma^{ij}\widehat{f_{,i}}\widehat{f_{,j}}+h(2g^{ij}\widehat{f_{,i}}%
\widehat{g_{,j}}-\tilde a^{ij}\cos(K_{m}x^{m})\widehat{f_{,i}}\widehat{f_{,j}%
})=0.  \label{e-1}
\end{equation}

In search for the solutions sufficing 
\begin{equation}
\left\{ 
\begin{array}{l}
\gamma^{ij}(y)\widehat{f_{,i}}\widehat{f_{,j}}=0 \\ 
2\gamma^{ij}\widehat{f_{,i}}\widehat{g_{,j}}-\tilde a^{ij}\cos(K_{m}x^{m})%
\widehat{f_{,i}}\widehat{f_{,j}}=0%
\end{array}
\right. .  \label{e-2}
\end{equation}
we get the following class of solutions: 
\begin{align}
\widehat{f} & =k_{i}(y)x^{i}+\Phi_{1}(y)  \label{e-3} \\
\ \ \gamma^{ij}k_{i}k_{j} & =0,
\end{align}
where $\Phi_{1}=\Phi_{1}(y)$ is arbitrary. Here $k_{i}$ is the wave 4-vector
of the EMW, $k_{1}=-\dfrac\omega c$, and $K_{m}$ is the wave 4-vector of the
GW. Considering $k^{i}=\gamma^{ij}k_{j}$, we get

\begin{equation}
\widehat{g}=\dfrac12\dfrac{\tilde a_{ij}k^{i}k^{j}}{K_{i}k^{i}}\sin\left(
K_{i}x^{i}\right) +\Phi_{2}(y,x);  \label{e-4}
\end{equation}

If we choose, for simplicity, $\Phi_{1}=\Phi_{2}=0,$ we obtain the eikonal 
\begin{equation}
\psi=k_{i}(y)x^{i}+h\tilde A(y)\sin\left( K_{i}x^{i}\right) ,~\ \ \gamma
^{ij}k_{i}k_{j}=0,~k_{i}=k_{i}(y)  \label{eikonal}
\end{equation}
where $\tilde A(y)=\dfrac12\dfrac{\tilde a_{ij}k^{i}k^{j}}{K_{i}k^{i}}%
=\dfrac12\dfrac{\tilde a^{ij}k_{i}k_{j}}{\gamma^{ij}K_{i}k_{j}}.$ In
anisotropic spaces, the components, $k_{i},$ generally also depend on the
directional variables $y^{i}.$

\subsection{Generalized geodesics}

The Finslerian function, $F,$ corresponding to the deformed locally
Minkowskian metric, namely, $F^{2}=(\gamma_{hl}(y)+%
\varepsilon_{hl}(x,y))y^{h}y^{l}$, leads to the Euler-Lagrange equations

\begin{equation}
\dfrac{\partial F^{2}}{\partial x^{i}}-\dfrac{d}{ds}(\dfrac{\partial F^{2}}{%
\partial y^{i}})=0,  \label{E-L}
\end{equation}
that are equivalent to 
\begin{equation}
g_{ij}^{\ast}\dfrac{dy^{j}}{ds}+\dfrac{1}{2}(\dfrac{\partial^{2}F^{2}}{%
\partial y^{i}\partial x^{j}}y^{j}-\dfrac{\partial F^{2}}{\partial x^{i}})=0,
\label{E-L-2}
\end{equation}
where $s$ is the arclength $s={\overset{t}{\underset{0}{\int}}F(x(\tau
),y(\tau))d\tau} ,$ and $g_{ij}^{\ast}=\dfrac{1}{2}\dfrac{\partial^{2}F^{2}}{%
\partial y^{i}\partial y^{j}}.$ Performing the computations, we get, in
linear approximation,

\begin{equation}
\dfrac{dy^{i}}{ds}+\gamma^{it}(\Gamma_{tls}+\dfrac12\frac{\partial
\varepsilon_{sl}}{\partial x^{j}\partial y^{t}}y^{j})y^{s}y^{l}=0
\label{g-2}
\end{equation}

This equation has a physical meaning. For the locally Minkowskian space with
small anisotropic deformation, the force potentials consist of two terms.
The second term in brackets, originating from the anisotropy of the
deformation, is associated with the velocity and provides an analogue to the
second term in the expression for the Lorentz force in electrodynemics. This
illustrates the ideas discussed in the end of \cite{sip-C6},\cite{sip-book}.

If $\varepsilon_{ij}(x,y)=h\tilde a_{ij}(y)\cos(K_{m}(y)x^{m})$, then,
performing the derivations, we obtain the geodesic equations:

\begin{equation}
\dfrac{dy^{i}}{ds}+hA^{i}(y)\sin(K_{m}x^{m})+hB_{p}^{i}(y)x^{p}%
\cos(K_{m}x^{m})=0  \label{geoaux}
\end{equation}
where the tensorial coefficients 
\begin{align}
A^{i} & =-\dfrac{1}{2}\gamma^{it}y^{l}y^{s}[y^{j}\dfrac{\partial(K_{j}\tilde{%
a}_{sl})}{\partial y^{t}}+(K_{s}\tilde{a}_{tl}+K_{l}\tilde{a}_{ts}-K_{t}%
\tilde{a}_{ls})].  \label{AB} \\
B_{p}^{i} & =-\dfrac{1}{2}\gamma^{it}y^{l}y^{s}y^{j}K_{j}\tilde{a}_{sl}%
\dfrac{\partial K_{p}}{\partial y^{t}}=-\dfrac{1}{2}\gamma^{it}K_{0}\tilde{a}%
_{00}\dfrac{\partial K_{p}}{\partial y^{t}}.  \notag
\end{align}
depend only on the directional variables $y^{i}.$ Here $\tilde{a}_{00}\equiv%
\tilde{a}_{nm}y^{n}y^{m}$ and $K_{0}\equiv K_{i}y^{i}$. In particular,%
\textrm{\ }if $K_{i}$\ are constant, then $B_{p}^{i}=0,$ $i=1\div4$ and the
equations of geodesics simplify. Solving \ref{geoaux}, one can find that
unit-speed geodesics of the perturbed metric $g_{ij}(x,y)=\gamma _{ij}(x,y)+h%
\tilde{a}_{ij}(y)\cos(K_{m}(y)x^{m})$ are described by 
\begin{equation}
x^{i}(s)=\alpha^{i}s+\beta^{i}-\dfrac{h}{2}\gamma^{it}\dfrac{\partial }{%
\partial y^{t}}\left( \dfrac{\tilde{a}_{00}}{K_{0}}\right) \sin(K_{m}x^{m})-%
\dfrac{hx^{p}}{2}\gamma^{it}\dfrac{\tilde{a}_{00}}{K_{0}}\dfrac{\partial
K_{p}}{\partial y^{t}}\cos(K_{m}x^{m}),  \label{gengeod}
\end{equation}
where $\alpha^{i}$ and $\beta^{i}$ depend on the initial conditions. In
particular, if $K_{m}$\ are constant, geodesics of the perturbed metric obey 
\begin{equation}
x^{i}(s)=\alpha^{i}s+\beta^{i}-\dfrac{h}{2}\gamma^{it}\dfrac{\partial }{%
\partial y^{t}}\left( \dfrac{\tilde{a}_{00}}{K_{0}}\right) \sin(K_{m}x^{m}).
\label{k_const}
\end{equation}

From (\ref{gengeod}), we get that along geodesics $hx^{i}(s)\simeq
h(\alpha^{i}s+\beta^{i}),$ and $hy^{i}(s)=h\dfrac{dx^{i}}{ds}\simeq
h\alpha^{i}.$

\section{Weak anisotropic perturbation of the flat Minkowski metric}

In order to find explicitly the way in which the anisotropy could reveal
itself in observations, we will first search for those solutions which are
as close to the solutions given in Section 2 as possible. Let the initial
metric be the flat Minkowskian one $\gamma=diag(1,-1,-1,-1)$; then the
system (\ref{wavesol2}) becomes 
\begin{equation}
\left\{ 
\begin{array}{l}
\gamma^{hl}K_{h}K_{l}=0 \\ 
a_{~j}^{i}K_{i}=\dfrac{1}{2}a_{~i}^{i}K_{j} \\ 
(\dfrac{1}{2}a_{~i}^{i}K_{j})_{\cdot b}=0.%
\end{array}
\right.  \label{wavesol3}
\end{equation}

Let us choose $K_{1}=-K_{2}=\dfrac Dc,$ where\emph{\ }$D,c\in\mathbf{R}%
,~K_{3}=K_{4}=0,$ and $a_{3}^{3}=-a(y),~a_{4}^{4}=a(y),~\ a_{~j}^{i}=0~$for
all other $(i,j).$ Then (\ref{wavesol3}) is identically satisfied and the
perturbed Minkowski metric in the weakly anisotropic case can be expectedly
presented as 
\begin{equation}
g_{ij}=\left( 
\begin{array}{cccc}
1 & 0 & 0 & 0 \\ 
0 & -1 & 0 & 0 \\ 
0 & 0 & -1+a(y)\cos(\dfrac Dc(x^{1}-x^{2})) & 0 \\ 
0 & 0 & 0 & -1-a(y)\cos(\dfrac Dc(x^{1}-x^{2})%
\end{array}
\right) ,  \label{qMin}
\end{equation}
where $a(y)$ is an arbitrary scalar $0$-homogeneous function, small enough
such that $a^{2}\simeq0.$ When $a(y)$ is a constant, this metric reduces to
the perturbed Minkowski metric for the isotropic empty space and the results
of Section 2 are valid.

\subsection{Eikonal}

Let $a(y)$ in the (\ref{qMin}) be equal to $a(y)=h\widetilde{a}(y^{i})$,
where $\widetilde{a}(y^{i})$ is an arbitrary scalar $0$-homogeneous
function, $h^{2}\simeq0$. Then, $\tilde{A}=\dfrac{1}{2}\dfrac{\tilde{a}%
^{ij}k_{i}k_{j}}{\gamma^{ij}K_{i}k_{j}}=\dfrac{1}{2}\dfrac{c^{2}\widetilde{a}%
(y^{i})(k_{3}^{2}-k_{4}^{2})}{D(ck_{2}-\omega)},$ and the eikonal (\ref%
{eikonal}) takes the form 
\begin{equation}
\psi=-\dfrac{\omega}{c}x^{1}+k_{2}x^{2}+k_{3}x^{3}+k_{4}x^{4}+\dfrac{h}{2}%
\dfrac{c^{2}\widetilde{a}(y^{i})(k_{3}^{2}-k_{4}^{2})}{D(ck_{2}-\omega )}%
\sin\left( K_{i}x^{i}\right) ,  \label{eik-lm}
\end{equation}

\textbf{Example:}

Let $v\in X(M)$\ be an arbitrary vector field and $\tilde a=\dfrac{K_{i}y^{i}%
}{\gamma_{ij}v^{i}y^{j}},$\ where $K$\ is the wave vector. Thus, $\tilde a$
is globally defined. In the given local frame, in which $K_{1}=\dfrac Dc,$\ $%
K_{2}=-\dfrac Dc,$ $K_{3}=0,K_{4}=0$\ (which is, chosen such that the GW
propagates antiparallel to the $Ox$\ axis), $\tilde a$ is equal to $\dfrac{%
\dfrac Dc(y^{1}-y^{2})}{\gamma_{ij}v^{i}y^{j}}.$ In a simple case, $v^{i}=0,$
$i=1,2,3 $ and $v^{4}=-\dfrac Dc$, we get 
\begin{equation*}
\tilde a=\dfrac{y^{1}-y^{2}}{y^{4}}
\end{equation*}

hence,

\begin{equation}
\psi=-\dfrac{\omega}{c}x^{1}+k_{2}x^{2}+k_{3}x^{3}+k_{4}x^{4}+\dfrac{h}{2}%
\dfrac{c^{2}(y^{1}-y^{2})(k_{3}^{2}-k_{4}^{2})}{Dy^{4}(ck_{2}-\omega)}%
\sin\left( K_{i}x^{i}\right)  \label{eik-1}
\end{equation}

\subsection{Geodesics}

In the case of a small anisotropic perturbation of the Minkowski metric $%
diag(1,-1,-1,-1),$ geodesics are described by (\ref{k_const}): 
\begin{equation*}
x^{i}(s)=\alpha^{i}s+\beta^{i}-\dfrac{h}{2}\gamma^{it}\dfrac{\partial }{%
\partial y^{t}}\left( \dfrac{\tilde{a}_{00}}{K_{0}}\right) \sin(K_{m}x^{m}).
\end{equation*}
\ 

In particular, for $\tilde a=\tilde a(\dfrac{y^{1}-y^{2}}{\phi(y^{3},y^{4})}%
) $\ and $K_{1}=\dfrac Dc,\ K_{2}=-\dfrac Dc,~K_{3}=K_{4}=0,~\ D,c\in 
\mathbf{R},$ we substitute it into eq.(\ref{k_const}) and get $%
x^{1}-x^{2}=\alpha s+\beta;$ $y^{1}-y^{2}=\alpha\in\mathbf{R}$.
Particularly, the Cartesian $Oy$ coordinate is

\begin{equation}
x^{3}=\nu+u_{0}\left\{ \dfrac{x^{1}-x^{2}-\beta}\alpha+\dfrac{hc}{2D\alpha }%
\dfrac{\partial\tilde a_{00}}{\partial y^{3}}\sin(\dfrac
Dc(x^{1}-x^{2}))\right\} .  \label{x-3}
\end{equation}
where $u_{0},\nu\in\mathbf{R}.$

\textbf{Examples: }1)\textbf{\ }For $a=h=const$, where $\tilde{a}=1,$ we get
the expression obtained in \cite{sip-A} which immediately leads to (\ref%
{eq.8}).

2)\textbf{\ }If $a=h\dfrac{y^{1}-y^{2}}{y^{4}}$ as earlier, we get $a=\dfrac{%
h\alpha}{y^{4}}=\dfrac{1+ha_{1}}{y^{4}}$ along geodesics, and, with $%
x^{i}(0)=0,$ 
\begin{equation}
x^{3}=u_{0}\dfrac{x^{1}-x^{2}}{1+ha_{1}}+u_{0}\dfrac{hc}{y^{4}D}\sin(\dfrac
Dc(x^{1}-x^{2})).  \label{g-fM}
\end{equation}

Then the $Oy$-component of the atom velocity will contain a term of the form

\begin{equation*}
y^{3}\sim u_{0}\dfrac{hc}{y^{4}}\cos(\dfrac Dc(x^{1}-x^{2}))
\end{equation*}
and the amplitude factor in front of the cosine depends on the velocity
component, $y^{4},$ orthogonal to $Ox$ and $Oy$ axises.

\subsection{OMPR modification}

The physical interpretation of the obtained solutions leading to the
modifications in the OMPR effect is the following. One can see that the
anisotropy does not destroy the solution of the OMPR equations (\ref{eq.7}-%
\ref{eq.10}). For a simple anisotropic deformation of the Minkowski metric,\
we get the dependence of eqs. (\ref{eik-1}, \ref{g-fM}) on the directional
variable orthogonal to $Ox$\ and $Oy$, i.e. to the plane containing the
Earth, the space maser and the GW source.\emph{\ }

Geodesics describe the trajectory of the particle, and the sample eq. (\ref%
{g-fM}) means that the amplitude of the oscillations of the space maser atom
velocity component oriented at\emph{\ }the Earth, $y^{3}$, depends on $y^{4}$%
. This means that when the system \textquotedblright Earth-space maser-GW
source\textquotedblright\ is located close to the periphery of the galaxy,
the orientation of this system might affect the OMPR conditions. In our
example two of the OMPR conditions (\ref{eq.11}) must be modified and take
the form

\begin{align}
\frac{\alpha_{2}}{\alpha_{1}} & =\frac{\omega h}{8Dy^{4}}=b\varepsilon
;b=O(1);\varepsilon<<1  \label{cond-lM} \\
\frac{kv_{1}}{\alpha_{1}} & =\frac{\omega hc}{\alpha_{1}y^{4}}%
=\kappa\varepsilon;\kappa=O(1);\varepsilon<<1
\end{align}
that illustrates the qualitative analysis given in \cite{sip-book}. This
means that the experimental investigation of the astrophysical systems with
various orientations might provide the information on the quantitative
characteristics of the geometrical anisotropy (if any) of our galaxy.

\section{Weak perturbation of the anisotropic Berwald-Moor metric}

Instead of the anisotropic correction to the isotropic (Minkowski) metric,
we could try an originally anisotropic but still locally Minkowskian (i.e.
spatial variables independent) metric on\emph{\ }$\mathbf{R}^{4}$. Let us
consider the Finslerian Berwald-Moor metric $\gamma_{ij}(y)=\dfrac 12\dfrac{%
\partial^{2}F^{2}}{\partial y^{i}\partial y^{j}}$, in which $F=\sqrt[4]{%
y^{1}y^{2}y^{3}y^{4}}$. The explicit form of the unperturbed metric is
provided by the matrices

\begin{align}
\left( \gamma_{ij}\right) & =\left\{ 
\begin{array}{c}
-\dfrac{1}{8}\dfrac{F^{2}}{(y^{i})^{2}},~\ i=j\,\  \\ 
\dfrac{1}{8}\dfrac{F^{2}}{y^{i}y^{j}},~\ i\not =j%
\end{array}
\right. ;  \label{mBM} \\
\left( \gamma^{ij}\right) & =\left\{ 
\begin{array}{c}
-\dfrac{2}{F^{2}}(y^{i})^{2},~\ \ i=j, \\ 
\dfrac{2}{F^{2}}y^{i}y^{j},~\ \ i\not =j%
\end{array}
\right. .
\end{align}

The wave solutions, $\varepsilon_{ij}=a_{ij}(y)\cos(K_{m}x^{m}),$ for
Einstein's equations in vacuum for the anisotropic case are given by the
solutions of the system (\ref{wavesol2}), where the coefficients, $\overset{0%
}{C}\overset{}{_{~jd}^{i}}=\dfrac12\gamma^{ih}\dfrac{\partial\gamma_{hj}}{%
\partial y^{d}}$ will be given by 
\begin{equation}
\overset{0}{C}\overset{}{_{~jd}^{i}}=\dfrac p8\dfrac{y^{i}}{y^{j}y^{d}},~\
p=\left\{ 
\begin{array}{l}
-\dfrac38,\text{ \ \ \ if \ \ }i=j=d \\ 
\dfrac18,\text{ \ if\ }i=j\not =d\text{ \ or }i\not =j=d~\ \text{or \ }%
i=d\not =j. \\ 
-\dfrac18,~\ \ \ \ \ \ \ \ \text{if \ \ }i\not =j\not =d\not =i.%
\end{array}
\right.  \label{BM_Christoffel}
\end{equation}

If we choose the coordinate system such that 
\begin{equation}
K_{3}=K_{4}=0,\   \label{*}
\end{equation}
then the light-like condition $\gamma^{ij}K_{i}K_{j}$ leads to $K_{2}=\dfrac{%
y^{1}}{y^{2}}K_{1}.$ Moreover, $a_{ij}=h\lambda(y)K_{i}K_{j}$ (here $%
\lambda(y)$\ is an arbitrary scalar $0$-homogeneous function and $h$ is a
small constant $h^{2}\simeq0$) defines a solution of (\ref{wavesol2})
obeying the transverse traceless conditions $a_{~i}^{i}=0,~\
a_{~j}^{i}K_{i}=0.$ We got the following solution for the Einstein equations

\begin{equation}
\varepsilon_{ij}(x,y)=h\lambda K_{i}K_{j}\cos(K_{i}x^{i}),~\ i,j=1,...,4,
\label{E-BM-1}
\end{equation}
where the first component $K_{1}=K_{1}(y)$ of the wave vector is an
arbitrary $0$-homogeneous function of the directional variables.

Let us denote: $K_{1}=\dfrac D{y^{1}}$ and $n_{i}=\dfrac cDK_{i},~\
i=1,...,4.$ With these, the solution can be written as: 
\begin{equation}
\varepsilon_{ij}=h\dfrac{\lambda D^{2}}{c^{2}}n_{i}n_{j}\cos(\dfrac
Dc(n_{i}x^{i})).  \label{E-BM-2}
\end{equation}

In the given frame, we have $K_{2}=\dfrac{D}{y^{2}},$ $n_{3}=n_{4}=0.$

If, moreover, $x^{1}=ct$ and the preferred direction is $y^{i}=\dfrac{dx^{i}%
}{dt}=c\dfrac{dx^{i}}{dx^{1}}$ (the time derivatives of positional
variables), then $y^{1}=c,~\ n_{1}=1,~\ K_{1}=\dfrac Dc$ and the cosine in
the perturbation is $\cos(\dfrac Dc(x^{1}+n_{2}x^{2})).$

\textbf{Examples:}\emph{\ }

1) If $\lambda$ and $D$ are constant, $\ n_{i}=n_{i}(y),$ $i=1,2$ then 
\begin{equation}
\left( a_{ij}\right) =h\dfrac{\lambda D^{2}}{c^{2}}\left( 
\begin{array}{cccc}
(n_{1})^{2} & n_{1}n_{2} & 0 & 0 \\ 
n_{1}n_{2} & (n_{2})^{2} & 0 & 0 \\ 
0 & 0 & 0 & 0 \\ 
0 & 0 & 0 & 0%
\end{array}
\right) .  \label{D-const}
\end{equation}

2)\ A less simple example which is interesting because of its symmetry, is $%
\lambda D^{2}=y^{1}y^{2}.$ Then, 
\begin{equation}
\left( a_{ij}\right) =h\lambda D^{2}\left( 
\begin{array}{cccc}
\dfrac1{\left( y^{1}\right) ^{2}} & \dfrac1{y^{1}y^{2}} & 0 & 0 \\ 
\dfrac1{y^{1}y^{2}} & \dfrac1{\left( y^{2}\right) ^{2}} & 0 & 0 \\ 
0 & 0 & 0 & 0 \\ 
0 & 0 & 0 & 0%
\end{array}
\right) =h\left( 
\begin{array}{cccc}
\dfrac{n_{1}}{n_{2}} & 1 & 0 & 0 \\ 
1 & \dfrac{n_{2}}{n_{1}} & 0 & 0 \\ 
0 & 0 & 0 & 0 \\ 
0 & 0 & 0 & 0%
\end{array}
\right) ;  \label{a-gen}
\end{equation}

and we see that in this case the perturbation, $\varepsilon_{ij},$of the
metric becomes

\begin{equation}
\varepsilon_{ij}=h\dfrac{n_{i}n_{j}}{n_{1}n_{2}}\mathrm{\cos}[\dfrac
Dc(n_{1}x^{1}+n_{2}x^{2})],~\ i,j=1,2\emph{.}  \label{pert}
\end{equation}

\subsection{Eikonal equation}

With the help of $\tilde A=\dfrac12\dfrac{\tilde a^{ij}k_{i}k_{j}}{\gamma
^{ij}K_{i}k_{j}}=\dfrac12\dfrac{\lambda(K^{i}k_{i})^{2}}{K^{i}k_{i}}=\dfrac{%
\lambda(y)}2K^{i}k_{i},$\ eq. (\ref{eikonal}) yields the solution for the
eikonal (\ref{eikonal}). Rewriting $\tilde A=\dfrac{\lambda D}{2c}n_{i}k^{i}$%
, one obtains the solution for the eikonal as:

\begin{equation}
\psi=k_{i}(y)x^{i}+h\dfrac{\lambda D}{2c}n_{i}k^{i}\sin[\frac{D}{c}\left(
n_{1}x^{1}+n_{2}x^{2}\right) ]  \label{eik-BM-1}
\end{equation}

\textbf{Example:} if $\lambda D^{2}=y^{1}y^{2},$\ then\emph{\ } 
\begin{equation*}
\psi=k_{i}(y)x^{i}+h\frac{c}{2D}\frac{(n_{1}k^{1}+n_{2}k^{2})}{n_{1}n_{2}}%
\sin[\frac{D}{c}\left( n_{1}x^{1}+n_{2}x^{2}\right) ]
\end{equation*}

Equation (\ref{eik-BM-1}) describes the eikonal of the wave propagating in
the model anisotropic space-time with the Berwald-Moor metric perturbed by
the GW. In the Berwald-Moor case, the components $k_{i}(y)$\ cannot be
constant, since the equation $\gamma^{ij}k_{i}k_{j}=0$\ does not have any
constant solutions (except the trivial one $k_{i}=0,i=1,...,4)$.

\subsection{Geodesics}

In the case under discussion we get $K_{0}=K_{1}y^{1}+K_{2}y^{2}=2D,\ \tilde
a_{00}=\lambda K_{0}^{2}=4\lambda D^{2},~\dfrac{\tilde a_{00}}{K_{0}}%
=2\lambda D$, and the unit-speed geodesics $(F=1)$\ in eq.(\ref{gengeod})
obey 
\begin{equation}
x^{i}(s)=\alpha^{i}s+\beta^{i}-h\gamma^{ij}\dfrac{\partial(\lambda D)}{%
\partial y^{j}}\sin(K_{m}x^{m})-h\gamma^{ij}\dfrac{\partial K_{p}}{\partial
y^{j}}x^{p}\lambda D\cos(K_{m}x^{m}).  \label{geo-BM-1}
\end{equation}

The simplest solutions are obtained for $D=D(y^{1},y^{2});$ in this case,
performing the calculations, we find $s=\dfrac{K_{1}x^{1}+K_{2}x^{2}}{%
K_{1}\alpha^{1}+K_{2}\alpha^{2}}$.\ Calculating the derivative and using the
initial conditions, $x^{i}(0)=0\Rightarrow\beta^{i}=0,$ $i=1,2,3,$ we find
that the last term in (\ref{geo-BM-1}) disappears for $y^{3},$ $y^{4}$ and
get 
\begin{equation}
y^{3}=\alpha^{3}-4\dfrac{h\lambda D^{2}y^{3}}{F^{2}}\cos(K_{m}x^{m}).
\label{y3-BM}
\end{equation}

\textbf{Example:}if $\lambda D^{2}=y^{1}y^{2},$ then

\begin{equation}
y^{3}=\alpha^{3}-4\dfrac{hF^{2}}{y^{4}}\cos(K_{m}x^{m})\overset{F^{2}=1}{=}%
\alpha^{3}-4h\dfrac1{y^{4}}\cos(K_{m}x^{m})  \label{y3-BM-2}
\end{equation}

\subsection{OMPR modification}

As in the previous case, the anisotropy does not destroy the OMPR effect
itself, but now the modifications are more pronounced. Eq.(\ref{eik-BM-1})
for the eikonal also gives a trichromatic EMW, but the amplitudes of the
sidebands and their frequencies are now different from the isotropic case.
The geodesics in the form (\ref{y3-BM}) shows that the amplitude of the
atomic oscillations is now also different. All this would affect the OMPR
conditions (\ref{eq.11}) and they would be modified in the following way

\begin{align}
\frac{\alpha_{2}}{\alpha_{1}} & =h\frac{\lambda D}{4c}n_{i}k^{i}=b%
\varepsilon;b=O(1);\varepsilon<<1  \label{cond-BM} \\
4h\frac\omega{\alpha_{1}}\frac{\lambda D^{2}}{c^{2}}\sqrt{\frac{%
n_{1}n_{2}n_{4}}{n_{3}}} & =\kappa\varepsilon;\kappa=O(1);\varepsilon<<1 \\
(\omega-\Omega+kv_{0})^{2}+4\alpha_{1}^{2} & =D^{2}n_{1}^{2}+O(\varepsilon
)\Rightarrow Dn_{1}\sim2\alpha_{1}
\end{align}

or, for the sample example, $\lambda D^{2}=y^{1}y^{2},$

\begin{align}
\frac{\alpha_{2}}{\alpha_{1}} & =h\frac c{4D}\frac{(n_{1}k^{1}+n_{2}k^{2})}{%
n_{1}n_{2}}=b\varepsilon;b=O(1);\varepsilon<<1  \label{cond-BM-s} \\
4h\frac\omega{\alpha_{1}}\sqrt{\frac{n_{4}}{n_{1}n_{2}n_{3}}} &
=\kappa\varepsilon;\kappa=O(1);\varepsilon<<1 \\
(\omega-\Omega+kv_{0})^{2}+4\alpha_{1}^{2} & =D^{2}n_{1}^{2}+O(\varepsilon
)\Rightarrow Dn_{1}\sim2\alpha_{1}
\end{align}

As in the previous Section, we find that the orientation of the system (see
Fig.1) would affect the observations. Calculating the left hand sides of the
second condition in (5.18) for the systems I and II, one can see that their
ratio is equal to the ratio of the star velocity corresponding to the
galactic rotation and the star velocity in the direction of the galaxy axis.
Therefore, if we take two equivalent astrophysical systems that initially
suffice the OMPR conditions and differ only by their orientation with regard
to the galactic plane, only one of them will produce an observable OMPR
signal.

\section{Discussion}

The main results obtained in this paper are the following. In search for the
modifications of the Einstein-Hilbert action due to the anisotropy of the
space-time, we have constructed two simple models of the anisotropic
space-time with the metrics containing small perturbations. The additional
terms lead to a change in the Einstein equations. We have shown that in the
anisotropic case Einstein equations for the empty space still have wave
solutions (gravitational waves), but now they become direction dependent and
their amplitudes, $a_{jk}(y),$ and wave numbers, $K_{j}(y),$ can become
coupled. We also performed the corresponding generalizations of the
equations for the eikonal and for the geodesics and used them to find how
the OMPR conditions would change in the anisotropic space-time. It turned
out that the orientation of the astrophysical system (taking part in the
OMPR) with regard to the galactic plane causes changes in the observable
effect, thus, giving one the possibility to experimentally investigate the
space-time geometrical properties on the galactic scale.

The expression for the ''simplest scalar''\ which can be used in the
variation principle based on the Einstein-Hilbert expression for the action
was particularized for our model. If the perturbed locally Minkowskian
metric can be presented as $g_{ij}(x,y)=\gamma_{ij}(y)+\varepsilon_{ij}(x,y)$%
, then the space-time anisotropy produces additional terms in Ricci tensor $%
\mathcal{R}_{~jk}$ which is to be calculated with regard to $%
\Gamma_{~jk}^{i} $ equal to

\begin{equation}
\Gamma_{~jk}^{i}=\dfrac{1}{2}\gamma^{il}(\dfrac{\partial\varepsilon_{lj}}{%
\partial x^{k}}+\dfrac{\partial\varepsilon_{lk}}{\partial x^{j}}-\dfrac{%
\partial\varepsilon_{jk}}{\partial x^{l}})=-\dfrac{1}{2}\gamma
^{il}(a_{lj}K_{k}+a_{lk}K_{j}-a_{jk}K_{l})\sin(K_{m}x^{m}),  \label{Chr-eps}
\end{equation}

It turned out that the generalized equations of geodesics for the
anisotropic space-time (\ref{g-2}) contain the ''force potentials''\
consisting of two terms. The second term is associated to the directional
variable (''velocity'') and provides an analogue to the corresponding term
in the expression for the Lorentz force in electrodynamics.

\section{Acknowledgements}

The work was supported by the RFBR grant No. 07-01-91681-RA\_a and by the
grant No.5 / 5.02.2008, between the Romanian Academy and Politehnica
University of Bucharest.

\end{document}